\documentstyle[eqsecnum, preprint, aps, fixes, epsfig]{revtex}
\tightenlines
\begin{document}

\preprint{UW/PT 98-9}

\title{Periodic Euclidean Solutions of $SU(2)$-Higgs Theory}
\author{Keith L. Frost and Laurence G. Yaffe} 
\address{University of Washington\\ Dept.~of Physics\\ 
Seattle, WA 98105-1560}
\date{\today} 
\maketitle

\begin{abstract}
We examine periodic, spherically symmetric, classical solutions of
$SU(2)$-Higgs theory in four-dimensional Euclidean space.  Classical
perturbation theory is used to construct periodic time-dependent
solutions in the neighborhood of the static sphaleron.  The behavior
of the action, as a function of period, changes character depending on
the value of the Higgs mass.  The required pattern of bifurcations of
solutions as a function of Higgs mass is examined, and implications
for the temperature dependence of the baryon number violation rate in
the Standard Model are discussed.
\end{abstract}

\section{Introduction}

Non-perturbative processes in quantum field theory such as tunneling,
decay of metastable states, and anomalous particle production, are
intimately connected with the existence and properties of non-trivial
solutions of the associated classical field equations.  In electroweak
theory, baryon number violating processes are a consequence of
topological transitions in which there is an order one change in the
Chern-Simons number of the $SU(2)$~gauge field.  At zero temperature,
such transitions are quantum tunneling events, and the rate of these
transitions is directly related to the classical action of instanton
solutions of $SU(2)$~gauge theory~\cite{tHooft_76}.  At sufficiently
high temperatures,%
\footnote
    {But below the critical temperature (or cross-over)
    where ``broken'' electroweak symmetry is restored.}
the dominant mechanism for baryon number violation involves classical
thermally activated transitions over the potential energy barrier
separating inequivalent vacuum states.  The configuration
characterizing the top of the barrier is the static sphaleron solution
of $SU(2)$-Higgs theory~\cite{Klinkhamer&Manton_84}; the energy of
this solution controls the thermally-activated transition
rate~\cite{KuzminRubakov&Shaposhnikov_85,Arnold&McLerran_88}.

As briefly reviewed below, when one lowers the temperature from the
sphaleron dominated regime, the topological transition rate is related
to the action of periodic classical solutions of the Euclidean field
equations with a period~$\beta$ equal to the inverse temperature.
Very little detailed information is known, however, about the
properties of periodic classical solutions in electroweak theory.

In this paper, we examine periodic solutions in $SU(2)$-Higgs theory
(which represents the bosonic sector of electroweak theory in the
limit of vanishing weak mixing angle).  We focus exclusively on
classical solutions which are real in Euclidean space.  Classical
perturbation theory is used to construct periodic time-dependent
solutions in the neighborhood of the static sphaleron.  The variation
in the period as one changes the amplitude of oscillation away from
the sphaleron (or equivalently the turning point energy of the
solution) is found to change sign depending on the value of the Higgs
mass.  Unlike the situation in many simpler models, there is no
periodic classical solution which smoothly interpolates between
instantons at low temperature (long period) and the sphaleron at high
temperature (short period).  We argue that, depending on the value of
the Higgs mass, one or two different branches of periodic solutions
must exist, connected at a bifurcation point.  The topological
transition rate (in the semi-classical approximation) must show a
``kink'' as the temperature varies between zero and the electroweak
transition temperature.

The non-existence of arbitrarily long period instanton-antiinstanton
solutions in $SU(2)$-Higgs theory, and the consequent necessity for
abrupt changes in the topological transition rate, has been previously
noted
\cite{HabibMottola&Tinyakov_96,Kuznetsov&Tinyakov_97}.
This phenomenon also occurs in two-dimensional $O(3)$ non-linear sigma
models with soft symmetry breaking terms, which mimic many features of
$SU(2)$-Higgs theory.  Periodic classical solutions in softly-broken
$O(3)$~sigma models have been studied in detail by several authors~%
\cite{HabibMottola&Tinyakov_96,Kuznetsov&Tinyakov_97}.
We will argue that the behavior of periodic solutions found in the
sigma model mimics the situation in $SU(2)$-Higgs theory when the
Higgs mass is sufficiently small, but that for larger Higgs mass an
additional bifurcation emerges.

The plan of this paper is as follows.  Section II briefly discusses
classical solutions in a simple example of a one-dimensional periodic
potential, and shows how bifurcations in solutions appear as one
deforms the potential.  This provides a useful analogy for the later
discussion of changes in periodic $SU(2)$-Higgs solutions as the Higgs
mass is varied.  Section III defines our notation for classical
$SU(2)$-Higgs theory and its reduction to a $1{+}1$~dimensional theory
for spherically symmetric configurations, and summarizes the action of
the relevant discrete symmetries.  Periodic classical solutions
resembling small, widely separated chains of instantons and
antiinstantons are the subject of Section IV.  Such solutions exist
for periods small compared to the inverse mass scales of $SU(2)$-Higgs
theory, and may be perturbatively constructed starting from
superpositions of very small pure gauge instantons and antiinstantons.
Section V contains a general discussion of classical perturbation
theory for periodic solutions which are small deviations away from a
static solution, and section VI discusses the practical issues which
arise when applying this formalism to $SU(2)$-Higgs theory.  Numerical
results are the subject of Section VII, and the stability of periodic
solutions close to the sphaleron is examined in Section VIII\@.  The
final section discusses the implications of these results, combined
with the instanton-antiinstanton analysis, on the global structure of
periodic classical solutions and on the temperature dependence of the
(semi-classical) topological transition rate.

Before continuing, we briefly review the relation between finite
temperature transition rates and periodic Euclidean classical
solutions.  The partition function~$Z(\beta)$ for a quantum system at
non-zero temperature~$\beta^{-1}$ may be written as a Euclidean
functional integral over configurations which are periodic with
period~$\beta$,
\begin{equation}
Z(\beta) = \int_{q(0) = q(\beta)} [{\mathcal D}q] \> \exp(-S_E[q]),
\label{Z-def}
\end{equation}
where $S_E$~is the Euclidean action.  For a semi-classical treatment
of a weakly coupled theory, the classical action~$S_E \sim 1/g^2$,
where $g$~is the small coupling constant, and the functional integral
can be estimated by the method of steepest descent applied to the
minima of~$S_E$.  Non-perturbative transition rates may be shown to be
related to saddle-point expansions about extrema of~$S_E[q]$
possessing one negative mode~%
\cite{Coleman_77,Callan&Coleman_77,Affleck_81,KhlebnikovRubakov&Tinyakov_91}.
In particular, Euclidean periodic solutions with one negative mode
determine the rate of quantum tunneling events at finite temperature.
Generalizing the usage of Coleman~\cite{Coleman_77}, we will
generically refer to such classical solutions as ``bounces''.  To
exponential accuracy, the number of tunneling events per unit time per
unit volume, at a temperature~$\beta^{-1}$, scales as
\begin{equation}
\Gamma \sim \exp(-S_E(\beta)),
\end{equation}
where $S_E(\beta)$~is the Euclidean action of the bounce solution
with period~$\beta$.

\section{A One-Dimensional Example}

The behavior of solutions in a toy model of a one-dimensional periodic
potential will provide a useful analogy.%
\footnote
	{%
	Similar discussion of much of this material may be found in
	Refs.~\cite{HabibMottola&Tinyakov_96,Kuznetsov&Tinyakov_97}.
	} 
Consider the dynamics for a single degree of freedom defined by the
Euclidean action
\begin{equation}
S_E[q] \equiv \int dt \, \left( \frac{1}{2} \, \dot q^2 + V(q) \right),
\label{S-1d}
\end{equation}
where the potential~$V(q)$ will be specified momentarily.  Extrema of
this action have a conserved energy~$E = -\frac{1}{2} \dot{q}^2 +
V(q)$, and it is a straightforward matter of integration to solve for
periodic classical solutions.  In particular, the period~$\beta$ is
given by
\begin{equation}
\beta(E) = \oint \frac{dq}{\sqrt{2[V(q) - E]}},
\label{b-1d}
\end{equation}
where the integration takes place over a periodic trajectory with
turning points $q_{\mathrm tp}$ defined by $E = V(q_{\mathrm tp})$.
The Euclidean action of the classical solution with period $\beta$ is
\begin{equation}
S_E(\beta) = \beta E(\beta) + \oint dq \, \sqrt{2[V(q) - E(\beta)]},
\label{cS-1d}
\end{equation}
where~$E(\beta)$ is obtained by inverting the equation for the period
(\ref{b-1d}).

If one varies the classical action~$S$ of periodic solutions, as a
function of the period~$\beta$, the derivative~$dS/d\beta$ is minus
the corresponding Hamiltonian (see, for example,~\cite{Landau_Mech}),
from whence
\begin{equation}
\frac{dS_E(\beta)}{d\beta} = E(\beta).
\label{dS-db}
\end{equation}

\begin{figure}
\begin{center}
\setlength {\unitlength} {1cm}
\begin{picture}(0,0)
  \put(-1.5,4.0){$V(q)/E_S$}
  \put(5.75,-0.5){$q/\pi$}
\end{picture}
\epsfysize=7.5cm
\epsfxsize=12.0cm
\epsfbox{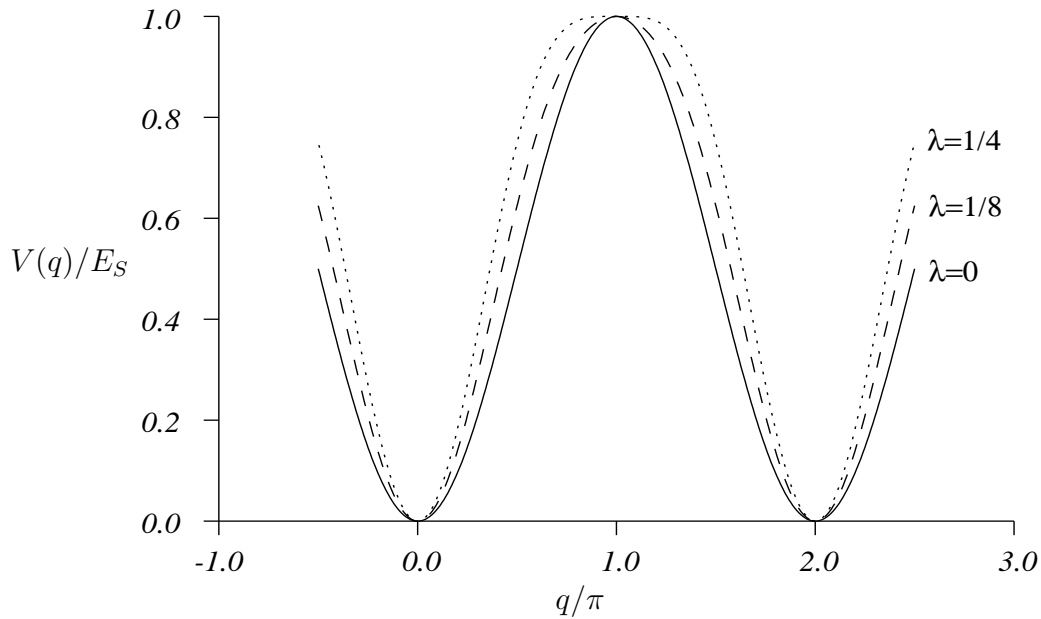}
\end{center}
\vspace{0.75cm}
\caption{The modified pendulum potential~$V(q)$~(\ref{V-1d}), plotted
for three values of the parameter~$\lambda$.  For~$\lambda=0$, it is
the sinusoid shown as the solid curve.  For~$\lambda = 1/8$, the
maximum at~$q=\pi$ is flattened, and the minimum at~$q=0$ sharpened,
as shown by the dashed curve.  This trend reaches its zenith
at~$\lambda = 1/4$, where, as shown by the dotted curve, the quadratic
term at the maximum vanishes.  Larger values of~$\lambda$ would cause
the maximum to split so that $V(q)$ would have two maxima per cycle.
\label{V-1d-plot}}
\end{figure}

Consider the potential of a modified pendulum
\begin{equation}
V(q) = \frac{E_S}{2} \left\{ [1{-}\cos(q)] + \lambda [1{-}\cos (2 q)]
\right\}, 
\label{V-1d}
\end{equation}
with~$0 \leq \lambda < 1/4$.  For~$\lambda = 0$, this reduces to a
sinusoidal potential, while larger values of~$\lambda$ flatten the
maxima and sharpen the minima of the potential.  Plots of the
potential for~$\lambda = 0$,~$1/8$,~and~$1/4$ are shown in
figure~\ref{V-1d-plot}.

To understand the branching behavior of bounce solutions in this
theory, one should examine the behavior of the potential~$V(q)$ near
the sphaleron at~$q=\pi$.  The Taylor expansion of the potential
about~$\pi$ begins
\begin{equation}
V(\pi{+}x) = \frac{E_S}{2}\left[2 - (1 {-} 4\lambda)\frac{x^2}{2!}
				+ (1 {-} 16\lambda)\frac{x^4}{4!} -
				{\mathcal O}(x^6) \right].
\label{V-1d-exp}
\end{equation}
In the inverted potential~$-V(q)$ in which Euclidean solutions evolve,
one finds harmonic motion for small perturbations about the single
sphaleron at~$q=\pi$ for~$\lambda < 1/4$.  For~$\lambda > 1/4$, as
noted above, the sphaleron splits in two and the solution at~$q=\pi$
becomes a local minimum of the action.

For~$\lambda < 1/16$, the quartic term in the
expansion~(\ref{V-1d-exp}) causes the inverted potential to soften
from what the simple harmonic term would give.  Oscillations thus
increase in period as their amplitude is increased, or alternatively,
as the turning point energy is lowered from the sphaleron energy.
This trend continues all the way to vanishing turning point energies,
where, at long periods, the bounce approaches a periodic kink plus
antikink.

For~$\lambda > 1/16$, however, one sees from the
expansion~(\ref{V-1d-exp}) that the quartic term agrees in sign with
the quadratic term in the potential, and thus causes the potential to
be steeper than what the simple harmonic term would give.  This
implies that the bounce, at least initially, decreases in period as
its amplitude is increased, or as the turning point energy is lowered
from the sphaleron energy.  At some critical turning point energy,
this trend must reverse in order to connect the bounce solution to the
kink-antikink at very long periods.

\begin{figure}
\begin{center}
\setlength {\unitlength} {1cm}
\begin{picture}(0,0)
  \put(-1.0,4.0){$S_E$}
  \put(5.75,-0.5){$\beta$}
\end{picture}
\epsfysize=7.5cm
\epsfxsize=12.0cm
\epsfbox{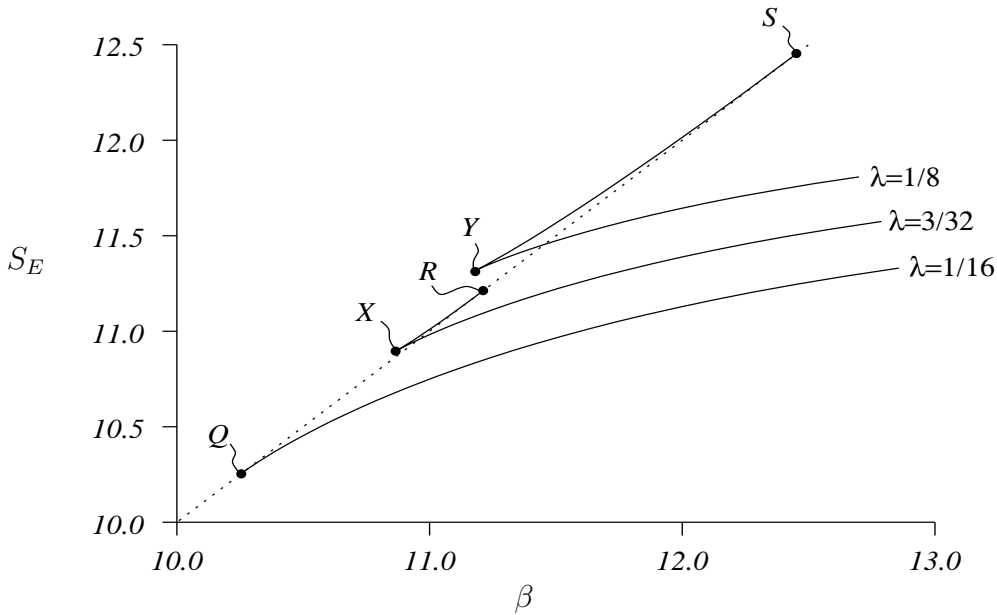}
\end{center}
\vspace{0.75cm}
\caption{Curves of action vs.\ period for periodic ``bounce''
solutions in the modified pendulum potential~(\ref{V-1d}), in units
where~$E_S = 1$.  The slope of a curve is the turning point energy.
The straight dotted line represents the static sphaleron.  The curve
merging with the sphaleron at point~$Q$ represents bounces with the
critical value of the parameter~$\lambda = 1/16$. Above this value
of~$\lambda$, the curves split at bifurcation points.  At~$\lambda =
3/32$, for example, the bounces pass from point~$R$ at the sphaleron,
towards shorter periods, until reaching point~$X$, beyond which
further decreases in the turning point energy take one towards longer
periods.  The branch~$RX$ grows longer for larger values of~$\lambda$,
becoming the branch~$SY$ for~$\lambda = 1/8$.
\label{pend-fig}}
\end{figure}

As noted above, this model is straightforward to integrate
numerically.  The results are shown in figure~\ref{pend-fig}, for
three different values of the parameter~$\lambda$.  The plot shows the
action of Euclidean classical solutions of the theory as functions of
the period.  As indicated by the derivative of the
action~(\ref{dS-db}), the slope of a curve on such a plot is the
turning point energy of the corresponding classical solution.  The
dotted line is the sphaleron solution, which, as a static
configuration, has the same energy at any value of the period.

At the critical value of~$\lambda = 1/16$, the bounce still increases
monotonically in period as its turning point energy is decreased,
giving the single curve of solutions starting from the sphaleron at
point~$Q$.  Above this value of~$\lambda$, the bounce initially
decreases in period as its turning point energy is decreased, then
reaches a bifurcation point, after which the bounce period begins to
increase again as the turning point energy is further decreased.
For~$\lambda = 3/32$, the bounce begins at the sphaleron at point $R$,
then proceeds to the bifurcation point~$X$ as the turning point energy
is lowered, after which further decreases in turning point energy take
one towards longer periods.  As~$\lambda$ is increased, the length of
the branch~$RX$ increases, becoming the branch~$SY$ at~$\lambda =
1/8$.

The (linearized) stability of a classical solution is determined by
the number of eigenmodes of the curvature of the action~$\delta^2 S$
(evaluated at the solution of interest) with negative eigenvalues.
Bifurcations, such as~$X$ or~$Y$ in figure~\ref{pend-fig}, correspond
to points in the space of solutions where a curvature eigenvalue
passes through zero.  When different branches of solutions merge at a
bifurcation such as~$Y$, the branch of solutions with larger action
has one more negative mode than the lower branch.  In the case of the
pendulum, the negative mode of the upper branch can be identified with
a deformation which moves the turning points further in or out, while
holding the period fixed.  Such a deformation causes no first order
change in action, since the momentum vanishes at the turning point.
But to second order, such a deformation can cause the action to
change.  The eigenvalue of this mode passes through zero at the
bifurcation point, and becomes positive on the lower branch of
solutions emerging from the bifurcation.

Points~$Q$, $R$, or~$S$ represent bifurcations where a branch of
periodic bounce solutions emerges from the static sphaleron.  Each
point on the branch~$SY$ (or~$RX$) represents a one-parameter family
of solutions related by time translations, {\em i.e.},~$x(t{-}t_0)$.
The time derivative~$\dot x(t{-}t_0)$ is a zero mode in the small
fluctuation spectrum and represents a deformation which is an
infinitesimal shift in~$t_0$.  The sphaleron, being static, has no
such zero mode.  At periods longer than that of the bifurcation~$S$
(or~$R$) the sphaleron has three negative modes: one is the static
unstable mode, corresponding to a deformation~$x \to x + \epsilon$
with time-independent~$\epsilon$, while the other two are periodic
oscillations,~$\delta x \propto \sin (2\pi t/\beta)$ or $\cos (2\pi
t/\beta)$.  At the bifurcation point, the two oscillating negative
modes become zero modes, and at shorter periods, positive modes of the
sphaleron.  On the branch~$SY$ (or~$RX$) of bounces emerging from the
sphaleron, one of the zero modes remains the time-translation zero
mode of the bounce, while the other becomes a negative mode --- the
same negative mode which later passes through zero at the
bifurcation~$Y$ (or~$X$) at the other end of this branch of solutions.
The bounce solutions also have one additional negative mode resembling
the static unstable mode of the sphaleron.

This connection between bifurcations and stability of solutions
will be important for the discussion in section IX\@.
Several figures illustrating the ``topography'' of the action
near bifurcations of the $SU(2)$-Higgs sphaleron,
and a more detailed discussion of the negative and zero modes
of the various solutions, appear in Section VIII\@.

At any temperature, the solution which characterizes the most probable
barrier crossings (and whose action controls the rate of barrier
crossing) is the bounce, or sphaleron, with the smallest action for
period~$\beta=1/T$.  As is clear from figure~\ref{pend-fig}, this
means that the branches, such as~$RX$ and~$SY$, which stay above the
action of the sphaleron, do not control the rate of barrier crossings
at any temperature.  Instead, for~$\lambda > 1/16$, there is an abrupt
change from sphaleron-dominated barrier crossing to bounce-dominated
crossing at the temperature where the curve of bounces crosses the
sphaleron line.  Physically, this is a transition between classical
thermally-activated transitions over the barrier, for temperatures
above the crossover, and quantum tunneling with a most-probable
energy~$E < E_S$ for temperatures below the crossover.%
\footnote{In the leading WKB approximation, the transition rate is
discontinuous at the crossover.  This discontinuity is smoothed into a
narrow transition region of width~${\mathcal O}(g^2)$ in the exact
transition rate.} 
In contrast, for~$\lambda < 1/16$, as one increases the temperature
from zero the most probable energy for tunneling grows until it
reaches~$E_S$, at which point the turning points of the most probable
tunneling path merge at the top of the barrier (and the WKB transition
rate smoothly interpolates between quantum tunneling and classical
thermal activation~\cite{Affleck_81}).

\section{The $SU(2)$-Higgs Model With Spherical Symmetry}

We now turn to the $SU(2)$~gauge theory in $3{+}1$~dimensions with a
single Higgs scalar in the fundamental representation.  This model
represents the bosonic sector of the Standard Model of the weak
interactions in the limit of small weak mixing angle.  The action may
be written in the form%
\footnote{For convenience, we have rescaled the gauge and Higgs fields
so that the action has an overall~$1/g^2$.  The conventional Higgs
vacuum expectation value is~$v=2M_W/g$, and the usual quartic coupling
is~$\lambda = \frac{1}{8}g^2 M_H^2 / M_W^2$.  As usual, the weak
fine-structure constant is~$\alpha_W = g^2/4\pi$.  In Minkowski space
(with a space-like metric), the usual action is minus
(\ref{action-4d}).}
\begin{equation}
  S = \frac{1}{g^2} \int d^4 x \left[ -\frac{1}{2} \mbox{Tr}
    (F_{\mu\nu}F^{\mu\nu}) + (D_\mu\Phi)^\dagger D^\mu\Phi + \frac{M_H^2}{8
      M_W^2}(\Phi^\dagger\Phi - 2M_W^2)^2 \right].
\label{action-4d}
\end{equation}
The scalar field~$\Phi$ is an $SU(2)$~doublet with covariant
derivative~$D_\mu = (\partial_\mu + A_\mu)$, where~$A_\mu \equiv A^a_\mu
\tau^a / 2 i$.  The gauge field strength is the commutator~$F_{\mu\nu}
= [D_\mu, D_\nu]$.

The action~(\ref{action-4d}) has an explicit $SU(2)_L$~gauge symmetry,
represented by $SU(2)$~matrices acting on the Higgs doublet~$\Phi$ (or
its conjugate~$\Phi^c \equiv -i\tau^2\Phi^*$) from the left.  It also
has a custodial~$SU(2)_R$ global symmetry, given by $SU(2)$ matrices
multiplying the matrix~$(\Phi, \Phi^c)$ from the right.

Spherically symmetric field configurations are those for which the
effect of a rotation can be undone by a gauge transformation combined
with a custodial~$SU(2)_R$ transformation.  Imposing spherical
symmetry reduces the four-dimensional theory to a two-dimensional
theory~\cite{Ratra&Yaffe_88,Yaffe_89}, which can be parameterized in
terms of six real fields~$\alpha$, $\beta$, $\mu$, $\nu$, $a_0$, and
$a_1$.  In terms of these two-dimensional fields, one may write the
original fields of the four-dimensional theory as
\begin{mathletters}
\begin{eqnarray}
  A_0 (\vec{r},t) & = & a_0 \, \hat{r} \cdot \vec{\tau} / 2 i\,,
  \\ 
  \vec{A}(\vec{r},t) & = &
  \frac{1}{2i} \left[ \frac{1}{r} (\alpha{-}1)\, \vec{\tau} \times \hat{r} +
    \frac{1}{r} \beta \,(\vec{\tau} - (\hat{r} \cdot \vec{\tau})\hat{r}) +
    a_1 (\vec{\tau} \cdot \hat{r}) \, \hat{r} \right] \label{defs-2d}, \\
  \Phi (\vec{r}, t) & = & (\mu + i \nu \vec{\tau} \cdot \hat{r}) \, \xi \,.
\end{eqnarray}%
\end{mathletters}
Here $\hat{r} \equiv \vec{r}/r$~is a unit radial vector, and $\xi$~is
a unit doublet which can be rotated arbitrarily by combined
global~$SU(2)_L$ and custodial~$SU(2)_R$ transformations.  It is
helpful to define complex linear combinations~$\chi \equiv \alpha + i
\beta$ and~$\phi \equiv \mu + i \nu$.  The field~$\chi$ then
represents the spherically symmetric degrees of freedom of the
tangential gauge fields, and~$\phi$ those of the Higgs field.

After imposing spherical symmetry in this form, the two-dimensional
theory which results has a $U(1)$~gauge symmetry, which is the
subgroup which remains of the original $SU(2)_L$~group.  As elements
of the original $SU(2)$~group, these $U(1)$~gauge transformations are
given by~$\Omega(\vec{r}, t) = \exp(i \omega(r,t) \vec{\tau} \cdot
\hat{r} / 2)$.  Under these Abelian gauge transformations,
\begin{mathletters}
\begin{eqnarray}
  \chi & \rightarrow & \exp(i \omega) \chi,  \\ 
  \phi & \rightarrow & \exp(i \omega / 2) \phi, \label{gauge-2d} \\ 
  a_\mu & \rightarrow & a_\mu + \partial_\mu \omega.
\end{eqnarray}%
\end{mathletters}
It is thus natural to define two-dimensional covariant derivatives as
\begin {mathletters}
\begin{eqnarray}
  D_\mu \chi & = & (\partial_\mu - i a_\mu) \, \chi, \\
  D_\mu \phi & = & (\partial_\mu - i a_\mu / 2) \, \phi \,.
\label{covd-2d}
\end{eqnarray}%
\end{mathletters}
Here, the reduced space-time indices~$\mu$ and~$\nu$ take on the
values~$0$~and~$1$, corresponding to the time~$t$ and distance from
the origin~$r$, respectively.  We also define a $(1{+}1)$-dimensional
$U(1)$~gauge field strength as~$f_{\mu\nu} \equiv \partial_\mu a_\nu -
\partial_\nu a_\mu$.

Substituting these definitions into the action (\ref{action-4d}) and
integrating out the angular coordinates gives for the two-dimensional
action
\begin{eqnarray}
  S & = & \frac{4 \pi}{g^2} \int dr \, dt \, \left[ \frac{1}{4} r^2
    f_{\mu\nu}f^{\mu\nu} + |D \chi|^2 + r^2 |D \phi|^2 +
    \frac{M_H^2}{8 M_W^2} \, r^2(|\phi|^2{-}2 M_W^2)^2 \right. \nonumber \\ 
  & & \left.  + \frac{1}{2 r^2}(|\chi|^2{-}1)^2
    + \frac{1}{2} |\phi|^2(|\chi|^2{+}1) - {\mathrm Re}(\chi^*\phi^2) 
	\right]\,.
\label{action-2d}
\end{eqnarray}
Note that finite action configurations must satisfy~$|\chi|\to 1$
as~$r\to 0$, and~$|\phi|\to \sqrt{2}M_W$,~$|\chi|\to 1$
with~$\chi^*\phi^2$ real and positive as~$r\to\infty$.

The transformations of the four- and two-dimensional fields under the
discrete symmetries of the $SU(2)$-Higgs model are summarized in
Table~\ref{CPT-table}.  When acting on spherically symmetric
configurations, four-dimensional charge conjugation can be undone by a
global $SU(2)_L$~rotation combined with a custodial
$SU(2)_R$~rotation, and therefore it has no effect on the fields of
the two-dimensional reduced theory.  Instead, $U(1)$~charge
conjugation in the two-dimensional theory is produced by a parity
transformation in four dimensions.

The sphaleron is a static field configuration for which the action is
stationary.  It may be found, in one choice of gauge, by imposing
$U(1)$~charge conjugation invariance on the spherically symmetric
fields defined above.  Referring to the action of parity in table
\ref{CPT-table}, one sees that $U(1)$ charge conjugation invariance
implies that~$a_\mu = 0$, and~$\phi$ and~$\chi$ are purely real.
There are thus only two real fields, $\alpha$~and~$\mu$.  The
sphaleron is the lowest energy configuration for which~$\alpha$
changes from~$-1$ to~$1$ as $r$~goes from zero to infinity.  The
static energy to be minimized can be written, by simplifying the above
action, as
\begin{equation}
E  =  \frac{4\pi}{g^2}\int dr \left[ 
	(\partial_r \alpha)^2 {+} r^2(\partial_r \mu)^2 + 
	\frac{1}{2 r^2}(\alpha^2{-}1)^2 + \frac{1}{2} (\alpha{-}1)^2 \mu^2 + 
	\frac{M_H^2}{8 M_W^2} \, r^2(\mu^2 {-} 2 M_W^2)^2 \right].
\end{equation}
Making an appropriate substitution on the radial variable $r$
to map the positive real line onto a finite interval,
and then discretizing the
above expression, one can readily solve numerically for the sphaleron
fields $\alpha_{\mathrm sph}$ and $\mu_{\mathrm
sph}$~\cite{Klinkhamer&Manton_84,Yaffe_89,Kunz&Brihaye_89}.

\begin{table}
\begin{center}
\setlength{\tabcolsep}{3.0\tabcolsep}
\begin{tabular}{|r||r|r|r|}
\hline
{\bf Field} & {\bf C \hspace*{0.8em}} & {\bf P \hspace*{1.3em}} & 
	{\bf T \hspace*{1.5em}} \\*[0.5ex]
\hline
$\Phi(\vec{r}, t)$ & $\Phi^*(\vec{r}, t)$ & 
	$\Phi(-\vec{r}, t)$ & $\Phi(\vec{r}, -t)$ \\
$\vec{A}(\vec{r},t)$ & $\vec{A}^*(\vec{r},t)$ & 
	$-\vec{A}(-\vec{r},t)$ & $\vec{A}(\vec{r}, -t)$ \\
$A_0(\vec{r}, t)$ & $A_0^*(\vec{r}, t)$ &
	$A_0(-\vec{r}, t)$ & $-A_0(\vec{r}, -t)$ \\
\hline
$\phi(r,t)$ & $\phi(r,t)$ & $\phi^*(r,t)$ & $\phi(r,-t)$ \\
$\chi(r,t)$ & $\chi(r,t)$ & $\chi^*(r,t)$ & $\chi(r,-t)$ \\
$a_1(r,t)$  & $a_1(r,t)$  & $-a_1(r,t)$   & $a_1(r,-t)$  \\
$a_0(r,t)$  & $a_0(r,t)$  & $-a_0(r,t)$   & $-a_0(r,-t)$ \\
\hline
\end{tabular}
\end{center}
\caption[Discrete Symmetry Transformations]{The transformations of the
various four- and two-dimensional fields under the action of the
discrete symmetries of the $SU(2)$-Higgs model. $C$ is charge
conjugation, $P$~is parity, and $T$~is time reversal.  Charge
conjugation~$C$ can be compensated by a global $SU(2)_L$~rotation
combined with a custodial $SU(2)_R$~rotation, and so becomes an
identity transformation on the two-dimensional fields. Parity~$P$
produces the effect of $U(1)$~charge conjugation on the
two-dimensional fields.
\label{CPT-table} }
\end{table}

\section{Instanton-Antiinstanton Solutions}

The $SU(2)$-Higgs model discussed in the preceding section does not
have exact instanton solutions, that is, finite action solutions on
unbounded Euclidean space.  Derrick's theorem (see for
example~\cite{Coleman_AoS}) shows that a reduction in the
four-dimensional length scale of any purported instanton will decrease
its action.  There are, however, approximate solutions resembling
instantons of the pure $SU(2)$~theory with characteristic sizes much
smaller than the inverse mass scales of the theory, namely~$M_W^{-1}$
or~$M_H^{-1}$~\cite{tHooft_76}.  The action of these configurations
decreases as the instanton shrinks (asymptotically approaching the
pure gauge value of~$8\pi^2/g^2$), but in all other directions the
action is minimized.  At zero temperature, these small instantons
dominate topological transitions in $SU(2)$-Higgs theory; quantum
fluctuations stabilize the size of the relevant
instantons~\cite{tHooft_76}.

\begin{figure}
\begin{center}
\epsfysize=6.0cm
\epsfxsize=12.0cm
\epsfbox{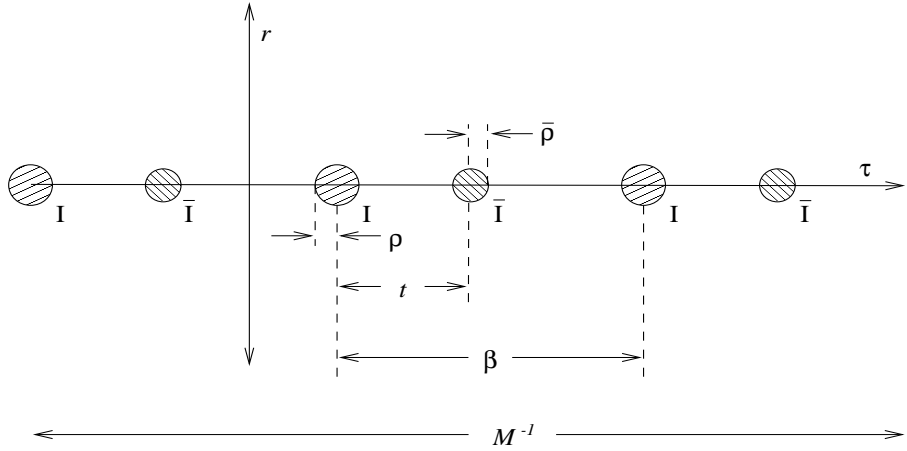}
\end{center}
\caption[Periodic Instanton-Antiinstantons]{A sketch of periodic
instanton-antiinstanton configurations.  The $r$~axis represents the
spatial distance from the origin, while the $\tau$~axis is imaginary
time.  The circles~$I$ are instantons, of size~$\rho$, while the
circles~$\bar{I}$ are antiinstantons, of size~$\bar{\rho}$.  The
distance from an instanton to the nearest antiinstanton is~$t$, while
the period is~$\beta$.  To construct such a solution iteratively,
$\rho$~and~$\bar{\rho}$~must be small compared to~$t$,~$\beta$,
or~$\beta{-}t$, which in turn must be much shorter than the inverse
mass scales of the theory, labeled here as $M^{-1}$.
\label{chain-fig}}
\end{figure}

Exact classical solutions resembling a periodic chain of small
instantons and antiinstantons also exist in $SU(2)$-Higgs theory.%
\footnote{Much of the following material may be found in
Ref.~\cite{KhlebnikovRubakov&Tinyakov_91}.}
Such a configuration is sketched in Figure~\ref{chain-fig}.  In order
for this configuration to be a classical solution, the attraction
between the instantons and antiinstantons must exactly balance the
tendency of the individual instantons or antiinstantons to collapse.
The resulting solutions may be constructed iteratively starting with a
superposition of $SU(2)$ pure gauge instantons, provided the instanton
sizes are small compared to their separation, and the separation is
small compared to the inverse mass scales~$M_H^{-1}$ or~$M_W^{-1}$.

To leading order, the fields of an instanton centered at the origin in
singular gauge are~\cite{tHooft_76}
\begin{eqnarray}
A_\mu(x) &=& \frac{2\rho^2}{x^2(\rho^2 + x^2)} \,
		\bar\eta^a_{\mu\nu}x_\nu (\tau^a / 2i)
		\> [1+{\mathcal O}(M_W x)], \nonumber \\
\Phi(x) &=& \frac{\sqrt {2} M_W \, x}{\sqrt{x^2+\rho^2}} \> \xi
		\> [1+{\mathcal O}(M_W x)+{\mathcal O}(M_H x)],
\label{inst-eq}
\end{eqnarray}
where $\xi$~is an arbitrary complex unit doublet, and
$\bar\eta^a_{\mu\nu}$~denotes 't Hooft's
$\eta$~symbol~\cite{tHooft_76}.
%
%
The action of this field configuration, to leading order in the small
parameter~$M \rho$ (where $M$ denotes either $M_W$ or~$M_H$) is
\begin{equation}
S_I = \frac{8\pi^2}{g^2} + \frac{4\pi^2}{g^2} M_W^2 \rho^2 +
	{\mathcal O}(M^4 \rho^4) \,.
\label{S-inst}
\end{equation}
The ${\mathcal O}(M_W^2 \rho^2)$~correction to the zero-order
instanton action is due to the derivative term for the Higgs field,
and is readily calculated as a surface integral~\cite{tHooft_76}.  The
neglected~${\mathcal O}(M^4 \rho^4)$ contribution comes from the Higgs
potential.

Consider placing such an instanton and antiinstanton along the time
axis a distance $t$ apart.  Since the fields approach their vacuum
values for large distances~$t \gg \rho$, one can construct an
approximate solution to the equations of motion by linear
superposition.  Furthermore, if~$\rho, \bar{\rho} \ll t \ll M^{-1}$,
the leading interaction term between an instanton and an antiinstanton
has the dipole-dipole form of the pure gauge
theory~\cite{CallanDashen&Gross_78}. It is
\begin{equation}
\Delta S_{I\bar{I}} = - \frac{1}{g^2} \left[ 
	\frac{96 \pi^2 \rho^2 \bar{\rho}^2} {t^4} + 
	{\mathcal O}(\rho^6 / t^6) + {\mathcal O}(M_W \rho^4 / t^3)
	\right] \,,
\label{dS-iibar}
\end{equation}
when the group orientations of the instanton and antiinstanton are
aligned to give the maximally attractive interaction.  This dipole
interaction arises from the pure gauge field part of the action, and
may be calculated as a simple surface integral.  The interaction term
for two instantons, or two antiinstantons, vanishes to this order.

For a periodic chain of alternating instantons and antiinstantons,
such as the one illustrated in figure~\ref{chain-fig}, there is a
dipole interaction between each instanton and antiinstanton.  Summing
over all of these interactions yields, for the action per period of
the periodic instanton-antiinstanton,
\begin{eqnarray}
S_{I\bar{I}}(\rho, \bar{\rho}, t; \beta) &=& \frac{4\pi^2}{g^2} \left[
	4 + M_W^2 (\rho^2{+}\bar\rho^2) -
	\frac{\rho^2 \bar\rho^2}{\beta^4} \sum_{n=-\infty}^{\infty} 
	\frac{24}{(n+t/\beta)^4} \right] \nonumber \\
	&=& \frac{4\pi^2}{g^2} \left[ 4 + M_W^2 (\rho^2{+}\bar\rho^2) - 
	\frac{8\pi^4 \rho^2 \bar\rho^2}{\beta^4} 
	\left( \frac{3{-}2\sin^2(\pi t/\beta)}
	{\sin^4(\pi t/\beta)} \right) \right]\,,
\label{S-perii0}
\end{eqnarray}
neglecting corrections of order $\rho^6/\beta^6$,~$M^4\rho^4$,
and~$M_W\rho^4/\beta^3$.  This expression is stationary with respect
to variations of~$t$ when~$t=\beta/2$.  Requiring the derivatives with
respect to~$\rho$ and~$\bar{\rho}$ to vanish then fixes
\begin{equation} 
\rho = \bar{\rho} = {\sqrt 2 M_W \beta^2 \over 4 \pi^2}\,,
\end{equation}
so that the classical action per period is 
\begin{equation} 
S(\beta) = {1 \over g^2}
\left[ 16\pi^2 + \frac{M_W^4 \beta^4}{2 \pi^2}
+ {\mathcal O}(M_W^5\beta^5) \right].
\end{equation}

This periodic solution has two negative modes.  One negative mode, due
to the attraction between the instanton and antiinstanton, is obvious
because the action~(\ref{S-perii0}) is maximized with respect to~$t$.
The other negative mode is a symmetric
perturbation~$\delta\rho=\delta\bar\rho$ in instanton and
antiinstanton sizes.  This can be easily verified by noting that the
mixed second derivative
\begin{equation}
\left. \frac{\partial^2 S}{\partial \rho \, \partial \bar{\rho}} 
	\right|_{\mathrm extremum} = - \frac{ 16 \pi^2 M_W^2}{g^2}
\label{perii-eig}
\end{equation}
is the only second derivative of~$S$ involving~$\rho$ or~$\bar{\rho}$
which doesn't vanish at the extremum of~$S$.  The right hand side of
the mixed second derivative~(\ref{perii-eig}) is thus a negative
eigenvalue of the curvature of~$S$, with an eigenvector
satisfying~$\delta\rho=\delta\bar{\rho}$ (and~$\delta t = 0$).  This
second negative mode is a signature of the delicate balancing required
to construct a genuine periodic solution out of widely separated
instantons, which themselves are not solutions.

The periodic instanton-antiinstanton solutions are time-reversal even,
and thus have turning points, where all of the time-reversal odd field
components (and velocities) vanish.  The turning points are located
halfway between the instanton and antiinstanton (at the origin in
figure~\ref{chain-fig}), as is clear by observing that the
time-reversed fields of an instanton are those of an antiinstanton,
and \emph{vice versa}.  The turning point energy of these solutions
vanishes as the period goes to zero.  Both negative modes of the
periodic instanton-antiinstanton solutions are parity odd and
time-reversal even.
	
\section{Thermal ``Bounces'' Near the Sphaleron}

The $SU(2)$-Higgs action~(\ref{action-4d}) is time-reversal invariant.
Solutions to the equations of motion must either be time-reversal
invariant, or come in time-reversed pairs.  The periodic bounces we
are interested in are time-reversal invariant solutions.  This means
that, at some time during the period, all time-reversal odd quantities
must vanish.  These times correspond to classical turning points, at
which the kinetic terms in the Lagrangian vanish, leaving just the
potential energy.

At turning-point energies just below the sphaleron energy, the bounces
will be small fluctuations away from the sphaleron.  Therefore, they
may be studied using classical time-dependent perturbation theory.
Given a time-independent action functional~$S[\phi]$ of fields
generically denoted as~$\phi$, and a static solution $\phi_0$ (the
sphaleron) which satisfies the equation of motion~$\delta S[\phi_0]
=0$, consider adding a small periodic fluctuation~$\delta\phi$, of
period~$\beta$, to~$\phi_0$.  The classical equation of motion can be
expanded in powers of the small fluctuation~$\delta\phi$:
\begin{eqnarray}
0 = \delta S[\phi] &=& \delta S[\phi_0 + \delta \phi]
    \nonumber \\ & = &
  S^{(2)} \cdot \delta \phi + \frac{1}{2!} \, S^{(3)} \cdot
  (\delta\phi)^2 + \frac{1}{3!} \, S^{(4)} \cdot (\delta\phi)^3 \,,
\label{pert-eom}
\end{eqnarray}
where the inner product implies summation over internal indices as
well as integration over space-time, and we have defined a shorthand
notation for variational derivatives of the action
\begin{equation}
S^{(n)} \equiv \delta^n S[\phi_0].
\label{dS-def}
\end{equation}

The leading order piece of the equation of motion (\ref{pert-eom}) is
\begin{equation}
0 = S^{(2)} \cdot \delta \phi  + {\mathcal O}\Bigl((\delta\phi)^2\Bigr).
\label{pert-eom1st}
\end{equation}
This implies that the curvature~$S^{(2)}$ has (at least) one
eigenvalue of magnitude ${\mathcal O}(\delta\phi)$; that is, which
becomes zero in the limit of vanishing fluctuation amplitude.  This
zero crossing of a curvature eigenvalue occurs at, and defines, the
critical period~$\beta_0$.  The presence of a zero mode at the
bifurcation point was pointed out in the context of the pendulum
model; here we see its necessity from perturbation theory.  We will
label this particular eigenvalue~$\lambda(\beta)$, explicitly
indicating its dependence on the period of the fluctuation, so that
\begin{equation}
    \lambda(\beta) = \lambda'(\beta_0) \, (\beta-\beta_0)
    + {\mathcal O}\Bigl((\beta{-}\beta_0)^2\Bigr) \,.
\label{lamb-exp}
\end{equation}
Let $\eta(\beta)$ denote the corresponding eigenmode, so that
\begin{equation}
S^{(2)} \cdot \eta(\beta) = \lambda(\beta) \, \eta(\beta) \,.
\label{eig-eq}
\end{equation}
To leading order, $\delta\phi$~consists of an undetermined (small)
amount of the zero mode~$\eta(\beta_0)$.

To solve for~$\delta\phi$ iteratively, beyond this linearized order,
we rewrite the equation of motion~(\ref{pert-eom}) using the
projection operator onto the eigenmode~$\eta$, denoted~$P_\eta$, to
separate the small eigenvalue~$\lambda$ when writing the inverse
curvature
 ${S^{(2)}}^{-1} = P_\eta/\lambda + {S^{(2)}}^{-1}(1{-}P_\eta)$.  This
gives
\begin{eqnarray}
\delta\phi & = & - {S^{(2)}}^{-1} \cdot \left( \frac{1}{2!} \, S^{(3)} \cdot
    (\delta\phi)^2 + \frac{1}{3!} \, S^{(4)} \cdot (\delta\phi)^3 \right)
    \nonumber \\
& = &
    \epsilon \, \eta - {S^{(2)}}^{-1} (1 - P_\eta) \cdot \left( \frac{1}{2!} \,
    S^{(3)} \cdot (\delta\phi)^2 + \frac{1}{3!} \, S^{(4)} \cdot
    (\delta\phi)^3 \right),
\label{pert-dphi}
\end{eqnarray}
where the second line of this expansion (\ref{pert-dphi}) follows from
the definition
\begin{equation}
\epsilon \equiv - \frac{1}{\lambda} \, \eta \cdot \left( \frac{1}{2!}\,
    S^{(3)} \cdot (\delta\phi)^2 + \frac{1}{3!}\, S^{(4)} \cdot (\delta\phi)^3
    \right) \,.
\label{eps-def}
\end{equation}
(For convenience, we have assumed that the eigenmode~$\eta(\beta)$ is
normalized to one.)  Defined in this way, $\epsilon$ is of magnitude
${\mathcal O}(\delta\phi)$, and will be used as a small parameter in
the expansion~(\ref{pert-dphi}).

The last line of the expansion~(\ref{pert-dphi}) defines a recursive
recipe for computing~$\delta\phi$ as a power series in the small
parameter~$\epsilon$.  Each place~$\delta\phi$ occurs in the last
line, one may substitute an entire copy of the last line, stopping at
whatever power of~$\epsilon$ is appropriate.  This expansion is most
conveniently written as a series of tree graphs, in which $n$-point
vertices (for $n=3$~and~$4$) correspond to~$-S^{(n)}$,
leaves~(---$\!\bullet$) emerging from any vertex represent factors
of~$\epsilon\eta$, and internal lines correspond to the
propagator~${S^{(2)}}^{-1}(1{-}P_\eta)$.  Each graph is to be divided
by the usual symmetry factor, given by the number of permutations of
leaves or vertices which leave the graph unchanged.

Carrying out the iterative expansion~(\ref{pert-dphi}) explicitly to
${\mathcal O}(\epsilon^3)$ gives us
\begin{equation}
\delta\phi = \; \raisebox{-0.22in}{\epsfig{file=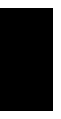}}\; +
	     \; \raisebox{-0.22in}{\epsfig{file=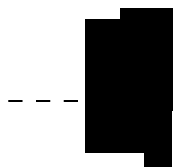}}\; +
	     \; \raisebox{-0.22in}{\epsfig{file=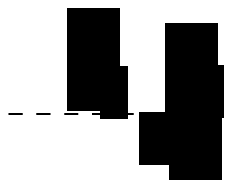}}\; +
	     \; \raisebox{-0.22in}{\epsfig{file=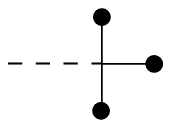}}\; + 
             \; {\mathcal O}(\epsilon^4)\,.
\label{diag-phi3}
\end{equation}

The graphical expansion~(\ref{diag-phi3}) shows the beginning of the
perturbative series for the field fluctuation~$\delta\phi$ in the
small parameter~$\epsilon$ which may be used to characterize its size.
But the parameter~$\epsilon$ is not independent of the oscillation
period~$\beta$.  The defining equation~(\ref{eps-def}) relates the
value of~$\epsilon$ to the period $\beta$, and may be used to generate
a perturbative expansion of the period $\beta$ in powers
of~$\epsilon$.  To see how this goes, expand equation~\ref{eps-def}
using the same graphical expansion for~$\delta\phi$,
\begin{equation}
\epsilon^2 \lambda = 
	3 \left( 
	\;\raisebox{-0.22in}{\epsfig{file=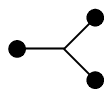}}\; 
	\right)\>
      + \> 4 \left( 
	  \;\raisebox{-0.22in}{\epsfig{file=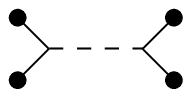}}\;
        + \;\raisebox{-0.22in}{\epsfig{file=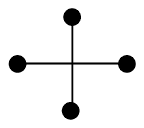}}\; 
	\right)\>
      + \> {\mathcal O}(\epsilon^5)\,.
\label{diag-lambda}
\end{equation}
In fact, all diagrams in the expansion~(\ref{diag-lambda}) with odd
powers of~$\epsilon$ (and hence~$\eta$) vanish.  This is because these
diagrams imply integration over all time variables, and $\eta$ has a
definite frequency~$\beta^{-1}$, as will be seen explicitly in the
following section.  To fully expand the left-hand side of the
$\lambda$~equation~(\ref{diag-lambda}), use the
expansion~(\ref{lamb-exp}) for the eigenvalue~$\lambda(\beta)$
and then expand~$\beta{-}\beta_0$ in powers of~$\epsilon$.  Since, as
noted above, only even powers of~$\epsilon$ occur on the right-hand
side of Eq.~(\ref{diag-lambda}), the same must be true in the
expansion of the period $\beta$,
\begin{equation}
\beta - \beta_0 = \beta^{(1)}\epsilon^2 +
\frac{1}{2!}\,\beta^{(2)}\epsilon^4 + \frac{1}{3!}\,\beta^{(3)}\epsilon^6
+ \cdots
\label{beta-exp}
\end{equation}
Substitute the period expansion~(\ref{beta-exp}) into the eigenvalue
expansion~(\ref{lamb-exp}), substitute the result into the $\lambda$
equation~(\ref{diag-lambda}), and match corresponding powers
of~$\epsilon$.  The leading order result is
\begin{equation}
\epsilon^4 \, \lambda'(\beta_0) \, \beta^{(1)} = 4 \left(
	\raisebox{-0.22in}{\epsfig{file=feyn11.eps}}
	\;+\; \raisebox{-0.22in}{\epsfig{file=feyn12.eps}} 
	\right)\,,
\label{beta1-diag}
\end{equation}
or more explicitly,
\begin{equation}
\beta^{(1)} = \frac{1}{\lambda'(\beta_0)} \left[ \frac{1}{2} \left(
  S^{(3)} \cdot \eta^2 \right) \cdot {S^{(2)}}^{-1}(1{-}P_\eta) \cdot \left(
  S^{(3)} \cdot \eta^2 \right) - \frac{1}{3!}\, S^{(4)} \cdot \eta^4
  \right]\,,
\label{beta1-alge}
\end{equation}
where all quantities may now be evaluated at the critical period~$\beta_0$.

\section{Computation}

We will evaluate equation~(\ref{beta1-alge}) for perturbations about
the sphaleron in $SU(2)$-Higgs theory, in order to find the
leading-order dependence of the period of sphaleron oscillations on
the amplitude~$\epsilon$ of the fluctuating fields.  We will also
evaluate the leading order dependence of the turning point energy on
the amplitude~$\epsilon$ in order to re-express dependence
on~$\epsilon$ as dependence on the turning point energy.

To carry this out, one must discretize the radial dependence in the
action~(\ref{action-2d}), solve for the static sphaleron, determine
the critical period~$\beta_0$ and the leading order
fluctuation~$\eta(\beta_0)$ by finding the zero-crossing eigenmode of
the curvature~$S^{(2)}$ evaluated at the sphaleron, and then assemble
the various pieces needed to compute~(\ref{beta1-alge}).

Since we are looking for periodic solutions, the time dependence of
each of the fields appearing in the field equation~(\ref{diag-phi3})
may be expanded in a Fourier series.  Because~$S^{(2)}$ is a
time-independent linear operator, only the vertices~$S^{(3)}$
and~$S^{(4)}$ generate couplings between different harmonics.  For the
calculation of~$\beta^{(1)}$, only field components at frequencies
$0$,~$\beta_0^{-1}$, and~$2\beta_0^{-1}$ are needed.

Fixing a gauge is required, so as to remove zero modes of the
curvature due to the residual $U(1)$~gauge invariance.  There are two
obviously reasonable choices: radial gauge~($a_1 = 0$), or temporal
gauge~($a_0 = 0$).  We chose to work in radial gauge, in order to
avoid having to introduce link variables in the one-dimensional radial
lattice.%
\footnote{Temporal gauge has a small advantage of allowing one to
reduce the Lagrangian to the separable form~$(\partial_t \phi)^2 +
V(\phi)$, if~$\phi$ are suitably rescaled field variables.  This
simplifies the calculation of the critical period~$\beta_0$, as
$-(2\pi/\beta_0)^2$ will simply equal the negative eigenvalue of the
static curvature~$V^{(2)}(\phi)$.  However, this choice has the cost
of introducing the link variables $a_1$ into the discretization of all
spatial derivatives on the radial lattice, and requires an otherwise
unnecessary similarity transformation on the field variables to
obtain the separable form.}

Because the action, and the sphaleron in our choice of gauge, are
parity (or $U(1)$~charge conjugation) even, it is useful to separate
the fluctuating fields into parity even and odd components.  The first
order fluctuation~$\eta$ is entirely odd.  Each diagram in the field
equation~(\ref{diag-phi3}) then produces a contribution which is even
if the number of leaves is even, and odd if the number of leaves is
odd.  The curvature~$S^{(2)}$ is block-diagonal, and only the parity
even block is needed for~(\ref{beta1-alge}).  Since the zero
mode~$\eta(\beta_0)$ is parity odd, the projection
operator~$(1{-}P_\eta)$ has no effect on the parity even block of the
propagator.

Table~\ref{pert-table} shows the discrete symmetries and leading order
perturbative expansions of the real fields appearing with this choice
of gauge.  Note that all time derivatives appearing in the action, or
its variations, may be trivially computed analytically.

The spatial derivatives in the action~(\ref{action-2d}) were
discretized using a uniform sampling of the transformed radial
variable~\cite{Yaffe_89}
\begin{equation}
s = \ln \left[ \frac{1 + M_H r}{1 + M_W r} \right] \Big{/} \ln (M_H / M_W), 
\label{s-def}
\end{equation}
which maps the semi-infinite line~$0 \leq r < \infty$ onto the unit
interval~$0 \leq s < 1$.  Once the action is discretized, the
sphaleron field, and the variations of the action at the sphaleron,
can be computed as functions of the period.  

\begin{table}
\begin{center}
\setlength{\tabcolsep}{3.0\tabcolsep}
\begin{tabular}{|c||c|c||c|}
\hline
{\bf Field} & {\bf P} & {\bf T} & 
	{\bf Leading Order Terms} \\*[0.5ex]
\hline
$\alpha(r,t)$ & $+$ & $+$ & $\alpha_{\mathrm sph}(r) + \epsilon^2
\left[\alpha^{(0)}(r) + \alpha^{(2)}(r)\cos(2 \omega t)\right]$ \\
$\mu(r,t)$    & $+$ & $+$ & $\mu_{\mathrm sph}(r) + \epsilon^2 
\left[\mu^{(0)}(r) + \mu^{(2)}(r)\cos( 2 \omega t )\right]$ \\
$\beta(r,t)$  & $-$ & $+$ & $\epsilon\, \beta^{(1)}(r) \cos(\omega t)$ \\
$\nu(r,t)$    & $-$ & $+$ & $\epsilon\, \nu^{(1)}(r) \cos(\omega t)$ \\
$a_0(r,t)$    & $-$ & $-$ & $\epsilon\, a_0^{(1)}(r) \sin(\omega t)$ \\
\hline
\end{tabular}
\end{center}
\caption[Perturbative Fields]{Symmetry properties, and leading order
terms, of the field components for small oscillations about the
sphaleron.  $P$~is parity (or equivalently $U(1)$~charge conjugation),
$T$~is time reversal, and a $+$~or~$-$ indicates whether the field is
even or odd, respectively, under the symmetry.  The static sphaleron
fields are labeled $\alpha_{\mathrm sph}$~and~$\mu_{\mathrm sph}$.
The fundamental angular frequency~$\omega$ is~$2\pi/\beta_0$,
where~$\beta_0$ is the critical period. The zero
eigenmode~$\eta(\beta_0)$ is composed of the leading order terms for
the parity odd fields~$\beta$, $\nu$, and~$a_0$.  Note that only even
powers of~$\epsilon$ occur in the expansions for parity even fields,
and only odd powers of~$\epsilon$ appear in parity odd fields.
\label{pert-table}}
\end{table}

In particular, the curvature of the action is a matrix which, if
inverted at the critical period, with the zero mode projected out,
would give the propagator.  In practice, it is computationally
inefficient to actually calculate the matrix inverse.  Since the
action is local, discretizing the radial derivatives as
nearest-neighbor differences gives the curvature matrix a band
diagonal structure.  Band diagonal algorithms for finding eigenvalues
and solving linear equations have a computational cost which scales
only linearly with the matrix dimension.  An algorithm for band
diagonal matrices~\cite{Schwarz_68} was used to find the eigenvalues
of the curvature matrix, including~$\lambda(\beta)$.  Another band
diagonal algorithm~\cite{Martin&Wilkinson_67} was used to solve the
linear equations~$S^{(2)}\cdot z = \zeta$, with~$\zeta =
S^{(3)}\cdot\eta^2$.  The same linear band algorithm was employed to
find the sphaleron fields~$\mu_{\mathrm sph}$~and~$\alpha_{\mathrm
sph}$ by Newton's method iterations, and to generate the
eigenvector~$\eta(\beta)$ by inverse
iteration~\cite{Martin&Wilkinson_67}.

As noted above, the variation~$\beta^{(1)}$ of period with amplitude
(\ref{beta1-alge}) may be converted into the variation in the period
of the bounce with respect to the turning point energy, at the
sphaleron.  Explicitly,
\begin{equation}
\frac{\partial \beta}{\partial E} = \beta^{(1)}
    \left( {\partial E \over \partial\epsilon^2}\right)^{-1}.
\end{equation}
Since the kinetic terms of the Lagrangian vanish at the turning
points, the variation of the turning point energy with
amplitude~$\partial E / \partial \epsilon^2$ is computed as the inner
product
\begin{equation}
\frac{\partial E}{\partial \epsilon^2}(\beta_0) =
	\frac{1}{\beta_0} \, \eta_{\mathrm tp}(\beta_0) 
	\,\cdot\, S^{(2)} \,\cdot\, \eta_{\mathrm tp}(\beta_0) \,,
\end{equation}
where~$\eta_{\mathrm tp}(\beta_0)$ are static fields equal to the
turning point values of~$\eta(\beta_0)$.  Similarly, the derivative
with respect to period of the eigenvalue~$\lambda'(\beta_0)$ appearing
in the~$\beta^{(1)}$ equation (\ref{beta1-alge}) is computed as the
inner product
\begin{equation}
\lambda'(\beta_0) = \eta(\beta_0) \,\cdot\, 
	\frac{\partial S^{(2)}}{\partial\beta} 
	\,\cdot\, \eta(\beta_0) \;.
\end{equation}

\section{Numerical Results}

\begin{figure}
\begin{center}
\setlength {\unitlength} {1cm}
\begin{picture}(0,0)
  \put(-1.5,4.0){$\beta_0 M_W$}
  \put(5.75,-0.5){$M_H/M_W$}
\end{picture}
\epsfysize=7.5cm
\epsfxsize=12.0cm
\epsfbox{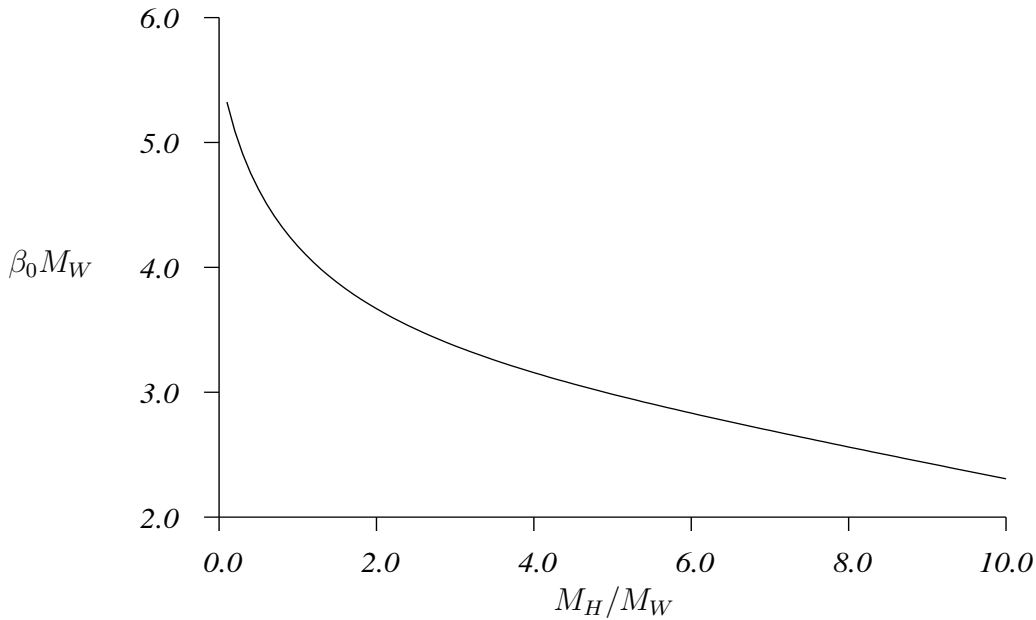}
\end{center}
\vspace{0.75cm}
\caption[Critical Period vs.\ Higgs Mass] {The critical period of
sphaleron oscillations~$\beta_0$ vs.\ the Higgs mass~$M_H$.
\label{beta-plot}}
\end{figure}

Figure~\ref{beta-plot} shows the dependence of the critical
period~$\beta_0$ of oscillations on the Higgs mass $M_H$.  The
critical period was found by adjusting an arbitrarily large
period~$\beta$ down until the eigenvalue of the oscillating negative
mode of the curvature matrix~$S^{(2)}$ crossed zero.

\begin{figure}
\begin{center}
\setlength {\unitlength} {1cm}
\begin{picture}(0,0)
  \put(-2.0,4.0){$\beta_0 M_{\mbox{sph}}\alpha_W$}
  \put(5.75,-0.5){$M_H/M_W$}
  \end{picture}
\epsfysize=7.5cm
\epsfxsize=12.0cm
\epsfbox{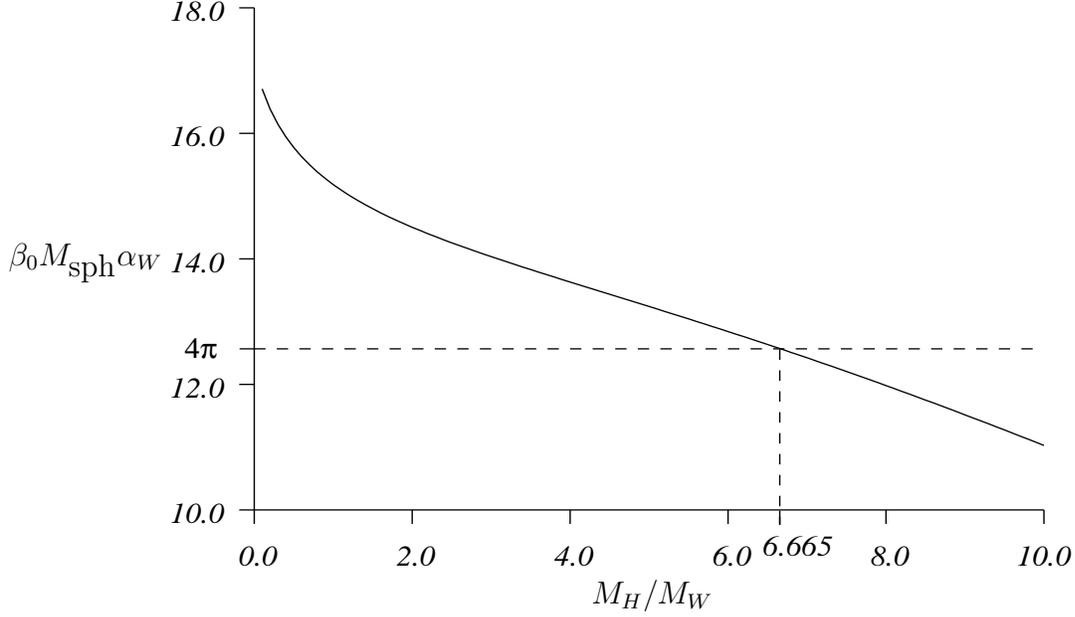}
\end{center}
\vspace{0.75cm}
\caption[Critical Period Sphaleron Action vs.\ Higgs Mass]{The action
of the sphaleron over a single critical period vs.\ the Higgs
mass~$M_H$.  The dashed horizontal line shows the minimum action of an
instanton-antiinstanton solution.
\label{S-plot}}
\end{figure}

Figure~\ref{S-plot} shows the action of the sphaleron for one critical
period~$\beta_0$ as a function of the Higgs mass~$M_H$. The sphaleron
action was calculated by simply multiplying the sphaleron energy by
the critical period~$\beta_0$.  This plot is important for
understanding the connection between the sphaleron oscillations and
the periodic instanton-antiinstanton solutions which occur at very
short periods.  The minimum action of these small
instanton-antiinstanton solutions over one period, which
is~$4\pi/\alpha_W$, is shown as the dashed horizontal line on the
graph.  The sphaleron action crosses this minimal
instanton-antiinstanton action when~$M_H \approx 6.665\, M_W$.

\begin{figure}
\begin{center}
\setlength {\unitlength} {1cm}
\begin{picture}(0,0)
  \put(-2.0,4.0){$\displaystyle\frac{\alpha_W}{M_W^2}
	\left(\frac{d\beta}{dE}\right)$}
  \put(5.75,-0.5){$M_H/M_W$}
\end{picture}
\epsfysize=7.5cm
\epsfxsize=12.0cm
\epsfbox{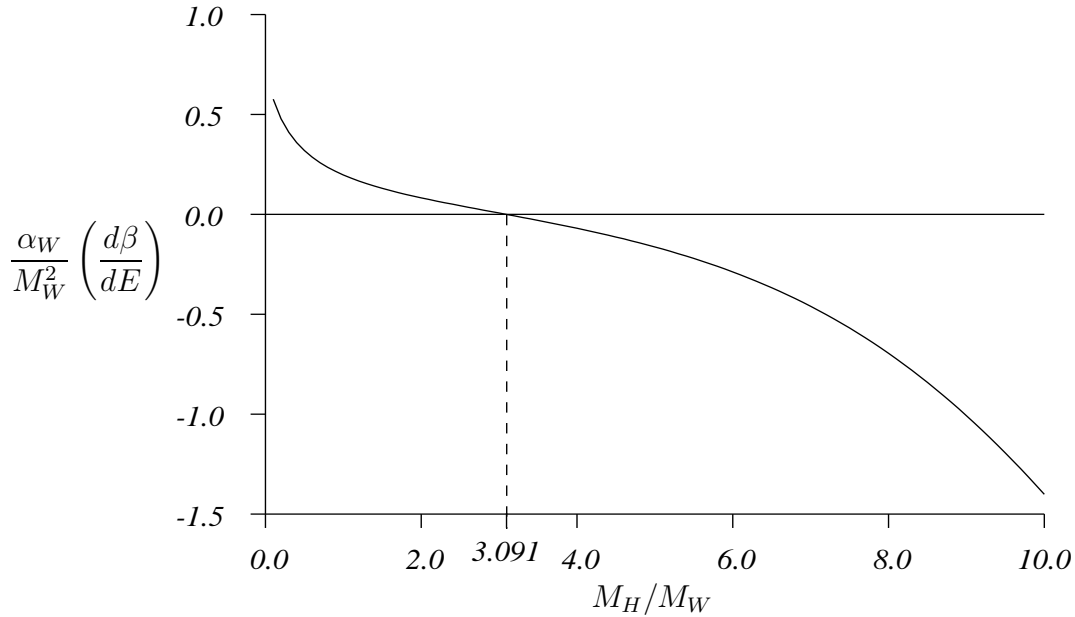}
\end{center}
\vspace{0.75cm}
\caption[$d\beta / dE$ vs.\ Higgs Mass]{The derivative of the period
of small sphaleron oscillations with respect to the turning point
energy vs.\ the Higgs mass~$M_H$.
\label{dbdE-plot}}
\end{figure}

Figure~\ref{dbdE-plot} shows the derivative of the period with respect
to the turning point energy, $\partial \beta / \partial E$, as a
function of the Higgs mass.  At low Higgs mass, the derivative is
positive, meaning that the oscillations move towards shorter periods
as the turning point energy is decreased.  At a Higgs mass of~$3.091\,
M_W$, the derivative crosses zero and thereafter becomes negative,
meaning that for Higgs masses larger than this value oscillations move
toward periods longer than~$\beta_0$ as the turning point energy is
initially decreased from the sphaleron energy.

\section{Stability}

Unstable modes of the bounce solution are closely related to those of
the sphaleron.  The sphaleron has a single static negative
mode~\cite{Yaffe_89}.  For periods shorter than the critical
period~$\beta_0$, this is its only negative mode.  For periods~$\beta$
longer than~$\beta_0$, the sphaleron has two additional negative modes
with frequency~$\omega = 2\pi/\beta$, given by the
eigenmode~$\eta(\beta)$ and its time derivative~$\partial_t
\eta(\beta)$.

\begin{figure}
\begin{center}
\setlength{\tabcolsep}{3.0\tabcolsep}
\begin{tabular}{c|c|c|}
&
{$\beta < \beta_0$} & {$\beta > \beta_0$} \\*[0.5ex]
\hline
{$\displaystyle\frac{\partial\beta}{\partial E}(\beta_0) > 0$ } &
\epsfxsize=4.0cm \epsfysize=4.0cm 
\vrule height 2.5cm depth 1.0cm width 0pt 
\smash{\raisebox{-1.5cm}{\epsfbox{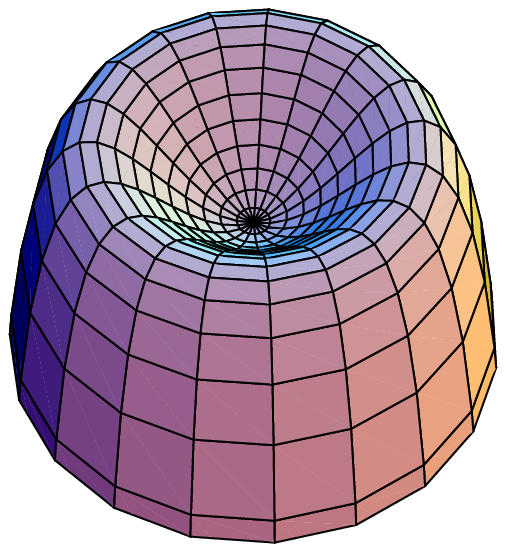}}} &  
\epsfxsize=4.0cm \epsfysize=4.0cm 
\smash{\raisebox{-1.5cm}{\epsfbox{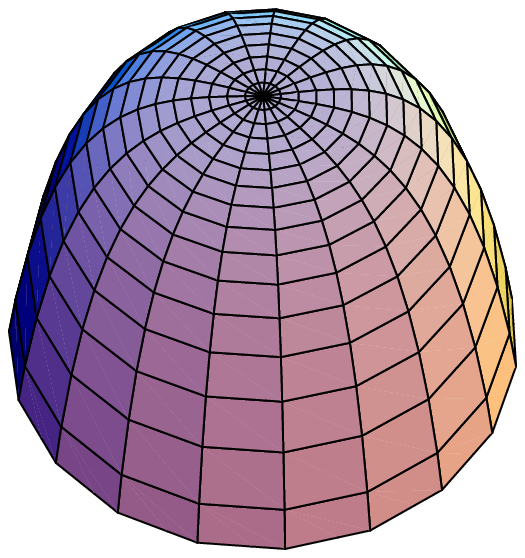}}} \\
& $\kappa_S = 1 \qquad \kappa_B = 2$ & $\kappa_S = 3$ \\
\hline
{$\displaystyle\frac{\partial\beta}{\partial E}(\beta_0) < 0$ }
&
\epsfxsize=4.0cm \epsfysize=4.0cm 
\vrule height 2.5cm depth 1.0cm width 0pt 
\smash{\raisebox{-2.0cm}{\epsfbox{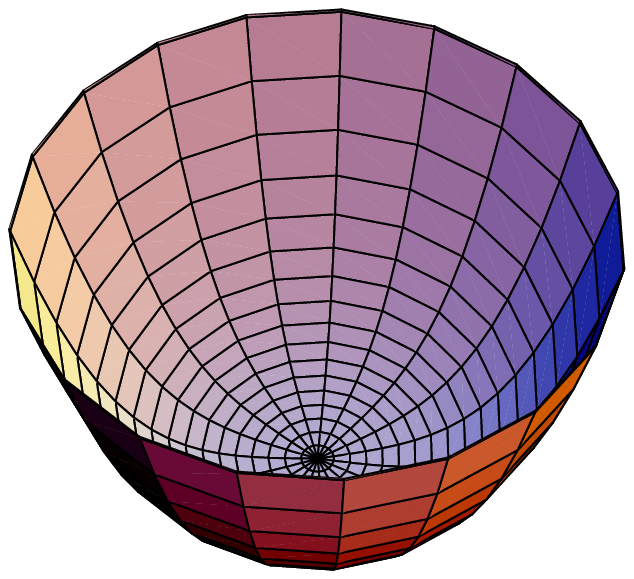}}} &
\epsfxsize=4.0cm \epsfysize=4.0cm 
\smash{\raisebox{-1.8cm}{\epsfbox{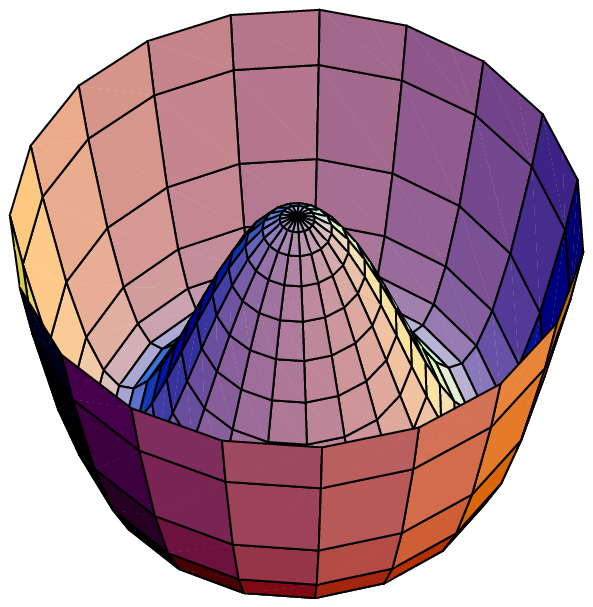}}} \\
& $\kappa_S = 1$ & $\kappa_S = 3 \qquad \kappa_B = 1$ \\
\hline
\end{tabular}
\end{center}
\caption[Action Topography]{Topography of the action near the
sphaleron and nearby bounce solutions.  The height of the surface in
each graph represents the action.  The extremum of the action on the
axis of rotational symmetry is the sphaleron.  The other extrema are
the periodic bounce solutions, and have a time translation zero mode
represented by rotations about the symmetry axis.  $\kappa_S$ denotes
the number of unstable modes of the sphaleron, while $\kappa_B$ is the
number of unstable modes of the bounce.  The action has one additional
unstable direction, not shown in the figure, corresponding to the
static unstable mode of the sphaleron.
\label{pic-table} }
\end{figure}

Figure \ref{pic-table} illustrates the landscape of the action near
the sphaleron and nearby bounce solutions, in the subspace spanned by
$\delta\phi$ and its time derivative $\partial_t \delta\phi$.  The
action is represented by the height of the surface in each graph.  The
periodic time translation symmetry is represented as the rotational
symmetry around the vertical axis.  Classical solutions appear as
peaks, cups, or saddle points.  The sphaleron sits on the symmetry
axis, because it is time translation invariant.  The direction
corresponding to the static negative mode of the sphaleron is not
shown.

As shown in figure \ref{pic-table}, when $\beta < \beta_0$, bounce
solutions appear as a circular ridge around the sphaleron and have two
negative modes (one of which is a small perturbation of the static
negative mode of the sphaleron, and is suppressed in figure
\ref{pic-table}), plus a time-translation zero mode.  When $\beta >
\beta_0$, bounce solutions appear as a circular valley around the
sphaleron and have only one near-static negative mode plus the
time-translation zero mode.  Both the near-static negative mode of the
bounce, and the additional negative mode for $\beta < \beta_0$ (which
is a small perturbation of $\eta(\beta_0)$) are parity odd and
time-reversal even.

\section{Discussion}

The results shown in figures \ref{S-plot} and \ref{dbdE-plot}, when
taken together with the analytic results on short-period periodic
instantons, allow one to suggest a picture of the branches of
periodic, spherically symmetric, Euclidean classical solutions of this
theory, which will be illustrated on the following plots of action
versus period.  On such plots, periodic classical solutions trace out
curves, the slope of which at any point is equal to the conserved
energy of the solution.  The sphaleron, as a static solution, appears
as a straight line through the origin, with a slope given by the
energy of the sphaleron.  Bounces with turning points have a slope
equal to their turning-point energy, and thus must always have a
positive slope.  Solutions may merge (at bifurcation points where
curvature eigenvalues pass through zero) when the curves representing
the solutions meet at tangent points on the plot.  The curves must be
tangent for the solutions to have the same conserved energy.

\begin{figure}
\begin{center}
\setlength {\unitlength} {1cm}
\begin{picture}(0,0)
  \put(3.0,-0.5){\footnotesize a) $M_H < 3.091\, M_W$}
\end{picture}
\epsfysize=5.0cm
\epsfxsize=8.0cm
\epsfbox{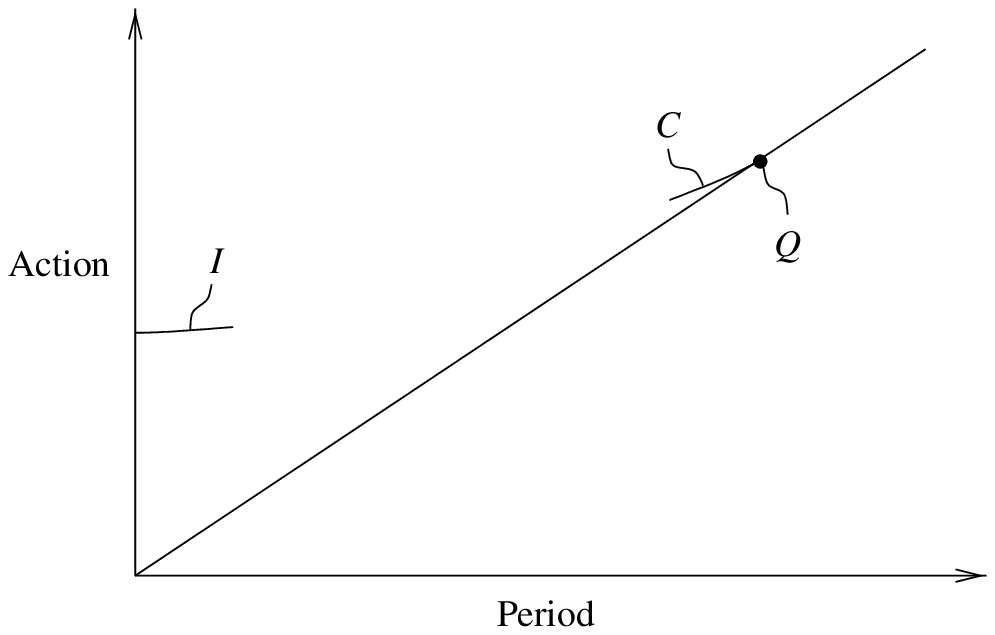}
\begin{picture}(0,0)
  \put(2.0,-0.5){\footnotesize b) $3.091\, M_W < M_H < 6.665\, M_W$}
\end{picture}
\epsfysize=5.0cm
\epsfxsize=8.0cm
\epsfbox{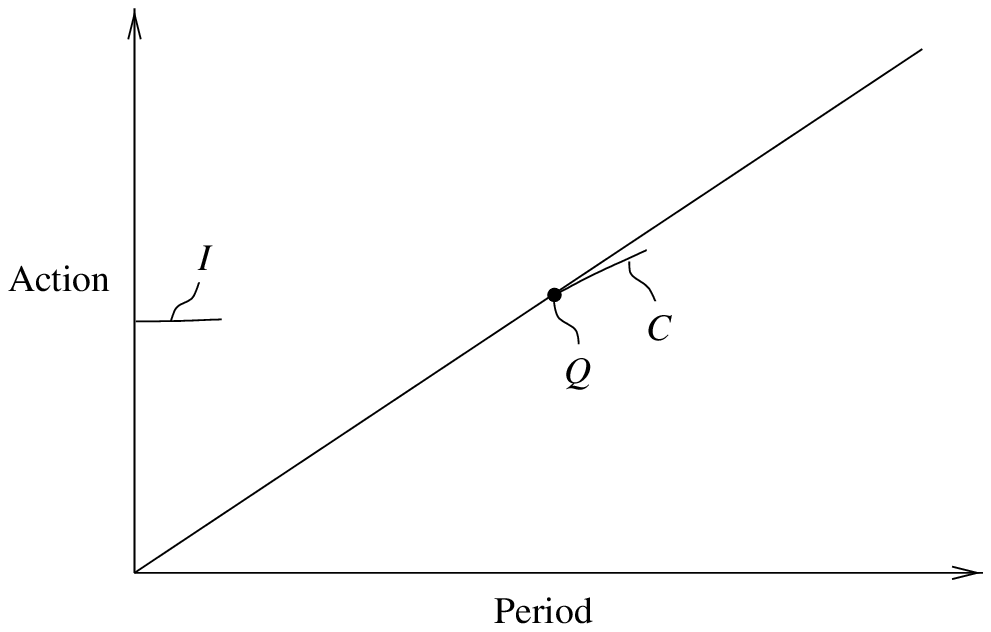}
\\[1.2cm]
\begin{picture}(0,0)
  \put(3.0,-0.5){\footnotesize c) $M_H > 6.665\, M_W$}
\end{picture}
\epsfysize=5.0cm
\epsfxsize=8.0cm
\epsfbox{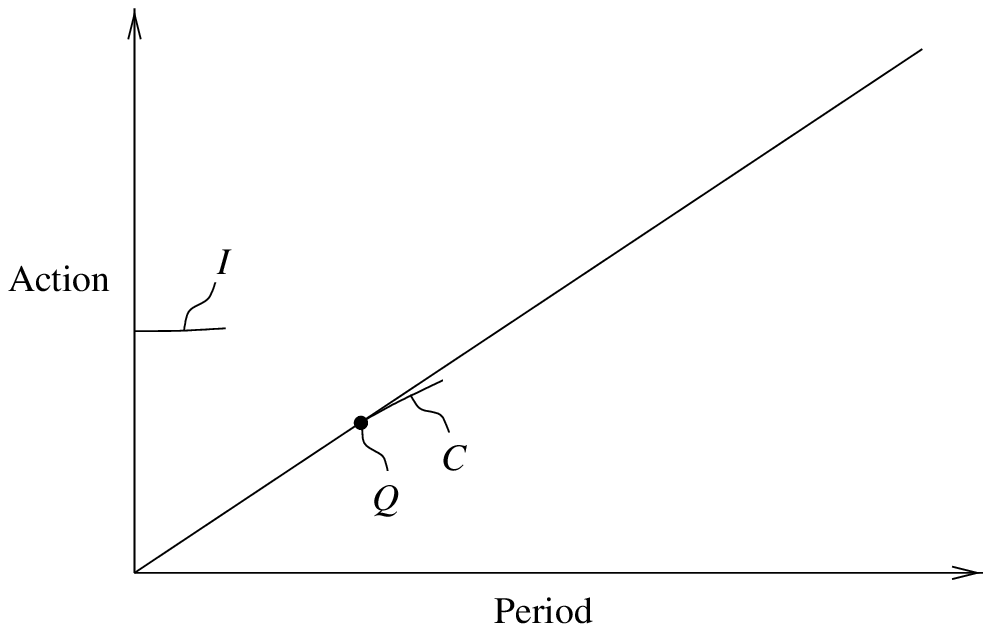}
\\[1.2cm]
\end{center}
\caption[Action vs.\ Period Results]{Euclidean classical solutions of
the $SU(2)$-Higgs model, sketched on plots of action versus period,
for three different values of the Higgs mass.  On these plots, slope
is equal to turning-point energy.  The plots are drawn for purposes of
illustration, and are not to scale.  The line through the origin is
the sphaleron.  The curve of bounces~$C$ merges with the sphaleron at
the point $Q$, which is at the critical period~$\beta_0$.  For periods
very short compared with the inverse mass scales
($M_H^{-1}$,~$M_W^{-1}$) of the theory, there exist periodic
instanton-antiinstanton solutions $I$ with near-zero turning point
energy.
\label{svb-known}}
\end{figure}

Figure \ref{svb-known} summarizes the perturbative results for the
bounces and periodic instanton-antiinstantons found in this paper, for
three different values of the Higgs mass.  These schematic plots of
action versus period are not drawn to scale.
The curve of bounces~$C$ merges with the sphaleron line at point~$Q$,
which is at the critical period~$\beta_0$.  For periods very short
compared with the inverse mass scales ($M_H^{-1}$,~$M_W^{-1}$) of the
theory, there exist periodic instanton-antiinstanton solutions~$I$
with near-zero turning point energy.  For brevity, we will henceforth
call these, in an abuse of language, periodic instantons.

For light Higgs (less than $3.091\, M_W$), figure~\ref{svb-known}(a)
shows that the bounces~$C$ decrease in period as their turning point
energy is decreased from the sphaleron energy.  This is clear from
figure~\ref{dbdE-plot} for small Higgs mass.  On the same plot, the
action at point~$Q$ is larger than the minimum action of the periodic
instantons~$I$.  This one can read from figure~\ref{S-plot} in the
limit of small Higgs mass.

As noted in figure~\ref{dbdE-plot}, above a Higgs mass of~$3.091\,
M_W$, the bounces~$C$ \emph{increase} in period as their turning point
energy is lowered from the sphaleron energy.  This situation is
illustrated in figure~\ref{svb-known}(b).

From figure~\ref{S-plot} we note that as the Higgs mass is increased,
the action at point~$Q$ decreases, because the period~$\beta_0$
shrinks faster than the sphaleron energy grows.  When the Higgs mass
reaches~$6.665\, M_W$, the action at point~$Q$ crosses below the
minimum action of the periodic instantons~$I$, that is,~$4 \pi /
\alpha_W$.  This situation is shown in figure~\ref{svb-known}(c).

Given this information available from perturbative calculations about
the branches of periodic instantons and bounces, one can construct a
likely scenario for the behavior of the various solutions in between
the known limits.

\begin{figure}
\begin{center}
\setlength {\unitlength} {1cm}
\begin{picture}(0,0)
  \put(3.0,-0.5){\footnotesize a) $M_H < 3.091\, M_W$}
\end{picture}
\epsfysize=5.0cm
\epsfxsize=8.0cm
\epsfbox{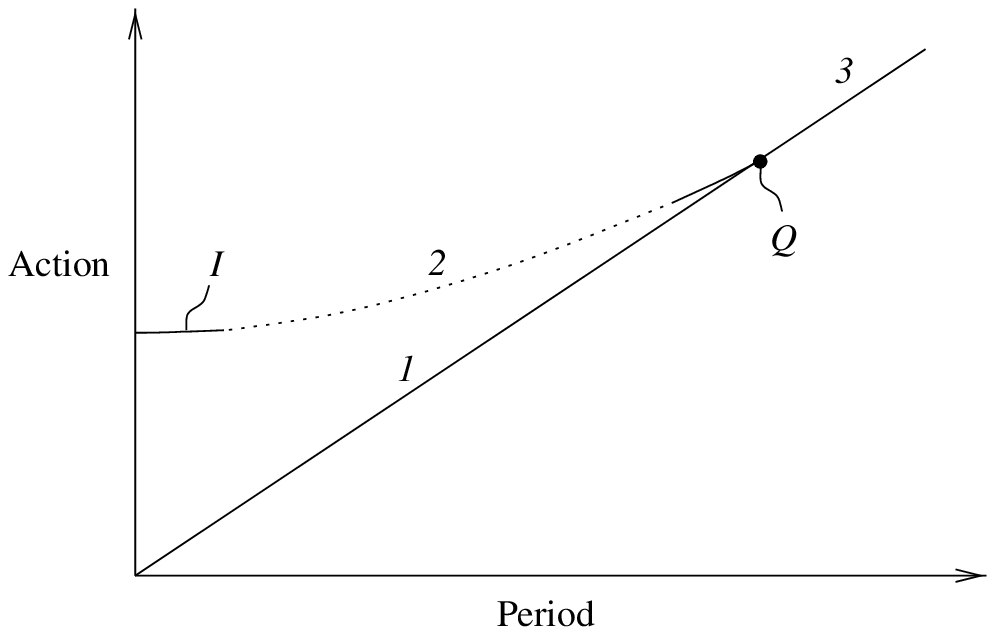}
\begin{picture}(0,0)
  \put(2.5,-0.5){\footnotesize b) $3.091\, M_W < M_H < 6.665\, M_W$}
\end{picture}
\epsfysize=5.0cm
\epsfxsize=8.0cm
\epsfbox{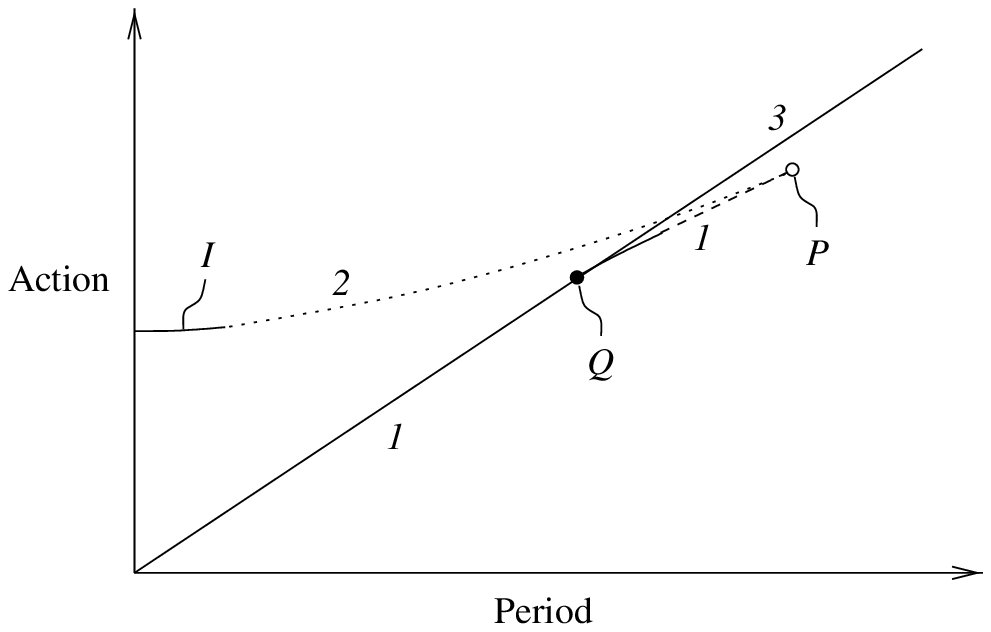}
\\[1.2cm]
\begin{picture}(0,0)
  \put(2.0,-0.5){\footnotesize c) $M_H > 6.665\, M_W$}
\end{picture}
\epsfysize=5.0cm
\epsfxsize=8.0cm
\epsfbox{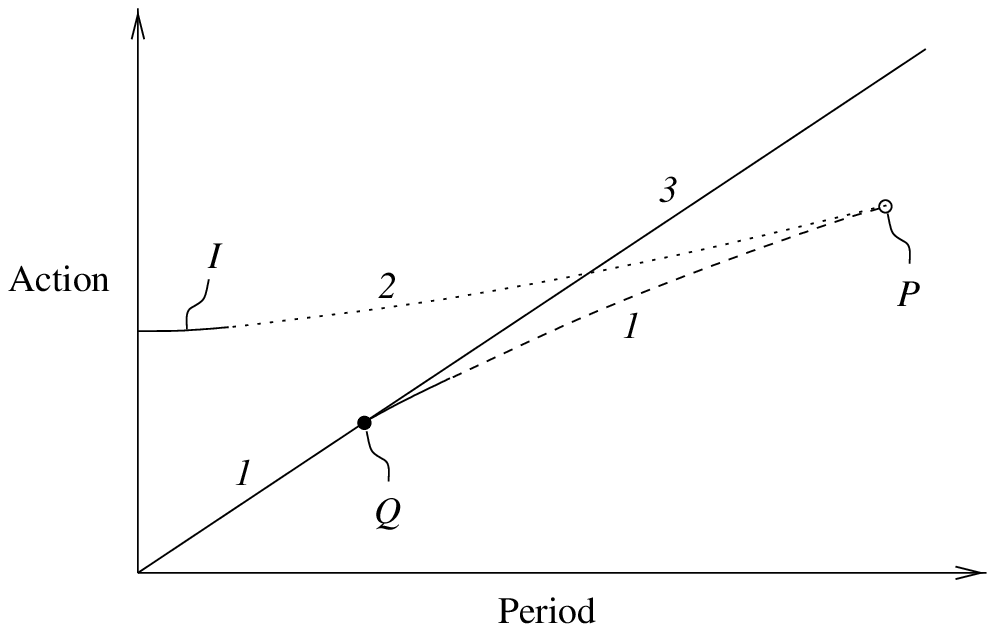}
\\[1.2cm]
\end{center}
\caption[Action vs.\ Period Hypotheses]{The simplest scenario for
Euclidean classical solutions of the $SU(2)$-Higgs model consistent
with the known perturbative results of figure~\ref{svb-known}.  The
number next to each branch of solutions gives the number of unstable
directions.  (a) For Higgs mass lighter than~$3.091\, M_W$, the
periodic instantons and bounces are joined by a single branch of
solutions, indicated here by a dotted line.  (b) For a Higgs mass
above~$3.091\, M_W$, the periodic instantons and bounces meet at a new
bifurcation point~$P$.  (c) When the Higgs mass exceeds~$6.665\, M_W$,
the action at point~$Q$ is less than the action of the periodic
instantons~$I$.
\label{svb-hypo}}
\end{figure}

For a sufficiently light Higgs mass, where the situation shown in
figure~\ref{svb-known}(a) holds, it is obviously simplest to assume
that the branch of periodic instanton solutions~$I$ and the branch of
bounces~$C$ are one and the same.  We illustrate this hypothesis in
figure~\ref{svb-hypo}(a), where the dotted line indicates the single
hypothetical branch of solutions.  Both of the solutions~$I$~and~$C$
have two parity odd, time-reversal even negative modes, which is an
important consistency check on this picture.  This picture of a single
branch of periodic solutions interpolating between the short-period
instanton-antiinstanton solution and the sphaleron at its critical
period is precisely the situation which occurs in the two-dimensional
non-linear $O(3)$~sigma model with a symmetry breaking mass term
\cite{HabibMottola&Tinyakov_96}.

When the Higgs mass exceeds the value~$3.091\, M_W$, as discussed
above, the bounce period increases as the turning point energy is
decreased from the sphaleron energy.  The simplest explanation for
this, drawn from analogy with the toy model, is shown in
figure~\ref{svb-hypo}(b). A new bifurcation point $P$ appears, where
the branch of bounces $C$ and the branch of periodic instantons $I$
merge, at a period longer than $\beta_0$.

When the Higgs mass reaches the value~$6.665\, M_W$, as noted above,
the action at point~$Q$ is equal to the minimum action of the periodic
instantons~$I$.  This situation is illustrated in
figure~\ref{svb-hypo}(c).  This implies that, to exponential accuracy,
the bounces set the rate of barrier penetration for the range of
periods for which their action is below the periodic instanton action.

As discussed earlier, bounces emerging from the sphaleron
with~$\beta>\beta_0$ have only a single negative mode (which is parity
odd and time-reversal even).  Hence the branch~$QP$ for~$M_H > 3.091\,
M_W$ will have only a single negative mode.  Point~$P$ is thus a
typical example of a bifurcation where real solutions with stabilities
differing by~$\pm 1$ merge.

To exponential accuracy, the rate of barrier penetration at a given
temperature~$\beta^{-1}$ in each of the plots in figure~\ref{svb-hypo}
is set by the lowest action configuration with period~$\beta$, except
that there is an additional lower limit set by the zero size limit of
periodic instanton configurations, which, for \emph{any} period, have
a limiting action of~$4\pi/\alpha_W$.  Thus, in cases~(a)~and~(b),
there is an abrupt crossover from a sphaleron-dominated rate at higher
temperatures to a singular instanton-dominated rate at lower
temperatures.  In case~(c), there is a smooth transition from the
sphaleron to the branch~$QP$, followed by an abrupt transition to
singular instantons.

\section*{Note Added}

The above scenario for periodic Euclidean solutions of $SU(2)$-Higgs
theory has recently been confirmed by direct numerical studies of
time-dependent (spherically symmetric) periodic solutions
\cite{Frost&Yaffe,HabibMottolaRebbiSingleton&Tinyakov}.

\section*{Acknowledgments}

The authors are grateful to S.~Habib, E.~Mottola, and R.~Singleton for
helpful discussions.  This work was supported in part by the U.S.\
Department of Energy grant DE-FG03-96ER40956.

\bibliographystyle{prsty}
\bibliography{perturb}

\end{document}